\documentclass[11pt,twoside]{article} \usepackage{amsfonts} 
\usepackage[mathscr]{eucal}
\usepackage{graphicx}
\usepackage{url}
\begin{document}

\title{\bf Quantum Measurement  \\ and the Paulian Idea} 
\author{\bf Christopher Fuchs and R\"udiger Schack} 
\date{}
\maketitle
\setcounter{page}{1}
\thispagestyle{empty}

%\begin{abstract}
%\smallskip

%\end{abstract}

%\section{Introduction}

%\vskip 0.5cm 

\begin{abstract}
\smallskip

In the quantum Bayesian (or QBist) conception of quantum theory, ``quantum measurement'' is understood not as a comparison of something pre-existent with a standard, but instead indicative of the creation of something new in the universe:  Namely, the fresh experience any agent receives upon taking an action on the world. We explore the implications of this for any would-be ontology underlying QBism.  The concept that presently stands out as a candidate ``material for our universe's composition'' is ``experience'' itself, or what John Wheeler called ``observer-participancy''.
\end{abstract}

\bigskip

{\small

\begin{quote}\raggedleft
{\it Of every would be describer of the
  universe one has a right to ask immediately two general questions.  The first
  is: ``What are the materials of your universe's composition?'' And the
  second: ``In what manner or manners do you represent them to be connected?''}

\smallskip

\hspace*{5.4cm} {William James 1988}

\end{quote}
}

\section{Introduction}

John Bell famously wrote that the word ``measurement'' should be banished from
fundamental discussions of quantum theory (Bell 1990). In this paper we
look at quantum measurement from the perspective of quantum Bayesianism, or
``QBism'' (Fuchs 2002a, 2004, 2010a, 2013, Caves {\it et al.}~2007), and argue that the word
``measurement'' is indeed problematic, even from our perspective. However, the reason it is problematic is not that the word is ``unprofessionally vague and ambiguous'', as Bell (1987) said. Rather, it is because the word's usage engenders a
misunderstanding of the subject matter of quantum theory.
%\footnote{For an argument in some sympathy with our own, but with a different 
%rhetorical flourish, see Mermin (2006).}
We say this because from the view of QBism quantum theory is a smaller theory than one might think -- it is smaller
because it indicates the world to be a bigger, more varied place than the
usual forms of the philosophy of science allow for. 

Crucial to the QBist conception of
measurement is the slogan -- inspired by Peres' (1978) more famous one --
that ``unperformed measurements have no outcomes''.
Mindful of James' injunction preceding this article, however, we believe that making precise the intuition behind
this slogan is the first step toward characterizing ``the materials
of our universe's composition''.

\section{Bayesian Probabilities}

Let us put quantum theory to the side for a moment, and consider instead basic
Bayesian probability theory (Savage 1954, de Finetti 1990, Bernardo and Smith 1994, Jeffrey 2004).
There the subject matter is an agent's
expectations for this and that.  For instance, an agent might write down a
joint probability distribution $P(h_i,d_j)$ for various mutually exclusive
hypotheses $h_i$, $i=1,\ldots,n$, and data values $d_j$, $j=1,\ldots,m$, appropriate to some phenomenon.  

A major role of Bayesian theory is that it
provides a scheme (Dutch-book coherence, see Vineberg 2011) for how these probabilities should be
related to other probabilities, $P(h_i)$ and $P(d_j)$ say, as well as to any
other degrees of belief the agent has for other phenomena.  The theory also
prescribes that if the agent is given a specific data value $d_j$, he should
update his expectations for everything else within his interest.  For instance,
under the right conditions (Diaconis and Zabell 1982, Skyrms 1987), he should reassess his
probabilities for the $h_i$ by conditionalizing:

\begin{equation}
P_{\rm new}(h_i) = \frac{P(h_i,d_j)}{P(d_j)}\;
\end{equation}

But what is this phrase
``given a specific data value''?  What does it really mean in detail?
Shouldn't one specify a mechanism or at least a chain of logical or
physical connectives for how the raw fact signified by $d_j$ comes into the
field of the agent's consciousness?  And who is this ``agent'' reassessing his
probabilities anyway?  Indeed, what is the precise definition of an agent? How
would one know one when one sees one?  Can a dog be an agent?  Or must it be a
person? Maybe it should be a person with a PhD?\footnote{This is a tongue-in-cheek
  reference to Bell (1990) again.}

We are thus led to ask: Should probability theory really be
held accountable for giving answers to all these questions?  In other words,
should a book like {\it The Foundations of Statistics} 
(Savage 1954) spend some of its pages demonstrating how the axioms of
probability -- by way of their own power -- give rise, at least in principle, to
agents and data acquisition itself?  Otherwise, should probability theory be
charged with being ``unprofessionally vague and ambiguous''?

Probability theory has no chance of answering these questions because they are
not questions within the subject matter of the theory.  Within probability
theory, the notions of ``agent'' and ``given a data value'' are primitive and
irreducible. Guiding agents' decisions based on
data is what the whole theory is constructed for -- just like primitive forces
and masses are what the whole theory of classical mechanics is
constructed for.  As such, agents and data are the highest elements within
the structure of probability theory -- they are not to be constructed from it,
but rather agents are there to receive the theory's guidance, and the
data are there to designate the world external to the agent.

\section{Quantum Bayesianism}

QBism says if all of this is true of Bayesian probability theory in general, it
is true of quantum theory as well.  As the foundations of probability theory
dismiss the questions of where data come from and what constitutes an agent,
so can the foundations of quantum theory dismiss them too.  

There will surely be a protest from some readers at this
point: ``It is one thing to say all this of probability theory, but
quantum theory is a wholly different story.'' Or: ``Quantum mechanics is no
simple branch of mathematics, be it probability or statistics.  Nor can it
plausibly be a theory about the insignificant specks of life in our vast
universe making gambles and decisions. Quantum mechanics is one of our best
theories of the world!  It is one of the best maps we have drawn yet of what is
actually out there.''
%\footnote{This is hardly conclusive evidence, but if one
%  does a Google search on the simultaneous terms ``quantum mechanics'' and
%  ``best theories of the world'' one will find about 22,000 hits.  If one does
%  a search on the terms ``quantum'' and ``best theories of the world'' one will
%  find about 36,000 hits.} 
 But this is where these readers err.
We hold fast: Quantum theory is simply {\it not} a ``theory of the world''.  Just like probability theory is not a theory
of the world, quantum theory is not as well. It is a theory for the use of
agents immersed in and interacting with a world of a particular character, the
quantum world.

By declaring this, we certainly do not want to dispense with the idea of a
world external to the agent.  Indeed it must be as Gardner (1983) says:
\begin{quote} \small
The hypothesis that there is an external world, not
dependent on human minds, made of {\it something}, is so obviously useful and
so strongly confirmed by experience down through the ages that we can say
without exaggerating that it is better confirmed than any other empirical
hypothesis.  So useful is the posit that it is almost impossible for anyone
except a madman or a professional metaphysician to comprehend a reason for
doubting it.
\end{quote}
Yet there is no implication in these words that quantum
theory, for all its success in chemistry, physical astronomy, laser making, and
so much else, must be read off as a theory of the world.  There is room for a
significantly more interesting form of dependence: Quantum theory is
conditioned by the character of the world, but yet is not a theory directly of
it.  Confusion on this point, we believe, is what has caused most of the
discomfort in quantum foundations in the 86 years since the theory's coming to
a relatively stable form.

\section{Measurement}

Returning to our discussion of Bell and the word ``measurement,'' it is not
because we think it unprofessionally vague and ambiguous that we regard
``measurement'' as problematic. It is because the word subliminally
whispers the philosophy of its birth -- that quantum mechanics {\it should} be
conceived in a way that makes no ultimate reference to agency, and that agents
are constructed out of the theory, rather than taken as the primitive
entities the theory is meant to aid.  In a nutshell, the word deviously carries
forward the impression that quantum mechanics should be viewed as a theory
directly of the world.

One gets a sense of the boundaries the word ``measure'' places upon our
interpretive thoughts by turning to any English dictionary.  Here is a sampling
from \url{dictionary.com/}:

\begin{itemize}
\vspace{-0.2cm}
\item to ascertain the extent, dimensions, quantity, capacity, etc., of,
esp. by comparison with a standard;
\vspace{-0.2cm}
\item to estimate the relative amount, value, etc., of, by comparison with some standard;
\vspace{-0.2cm}
\item to judge or appraise by comparison with something or someone else;
\vspace{-0.2cm}
\item to bring into comparison or competition.
\end{itemize}
\vspace{-0.2cm}
In not one of these definitions do we get an image of anything being created in
the measuring process; none give any inkling of the crucial contextuality of
quantum measurements, the context being a parameter ultimately set only in
terms of the agent.  Measurement, in its common usage, is something passive and
static: it is comparison between {\it existents}.  No wonder a slogan like
``unperformed measurements have no outcomes'' (cf.~Peres 1978) would seem irreparably
paradoxical.  If a quantum measurement is not comparison, but something else,
the only way out of the impasse is to understand what that something else is.

Correcting or modifying the word ``measurement'' is the prerequisite to a new ontology -- in other words,
prerequisite to a statement about the (hypothesized) character of the world
that does not make direct reference to our actions and gambles within it.
Therefore, as a start, let us rebuild quantum mechanics in terms more conducive
to the QBism program.
The best way to begin a more thoroughly delineation of quantum mechanics
is to start with two quotes on personalist Bayesianism itself.  The
first is from Hampton {\it et al.} (1973):
\begin{quote} \small
 Bruno de Finetti
believes there is no need to assume that the probability of some event has a
uniquely determinable value.  His philosophical view of probability is that it
expresses the feeling of an individual and cannot have meaning except in
relation to him.
\end{quote}
And the second is from Lindley (1982): 
\begin{quote} \small
The Bayesian, subjectivist, or coherent, paradigm is egocentric.  It is a tale
of one person contemplating the world and not wishing to be stupid
(technically, incoherent).  He realizes that to do this his statements of
uncertainty must be probabilistic.
\end{quote}

These two quotes make it
clear that personalist Bayesianism is a ``single-user theory''.  Thus, QBism
must inherit at least this much egocentrism in its view of quantum states
$\rho$.  The ``Paulian idea'' (Fuchs 2010b) -- which is also essential to the QBist view -- goes further still (cf.~Figure 1).  It says that the outcomes of quantum measurements are single-user as well!  That is to say, when an agent writes down her degrees of belief for the
outcomes of a quantum measurement, what she is writing down are her degrees of
belief about her potential {\it personal} experiences arising in consequence
of her actions upon the external world. 

\newpage

\begin{figure}
\includegraphics[height=6.7cm]{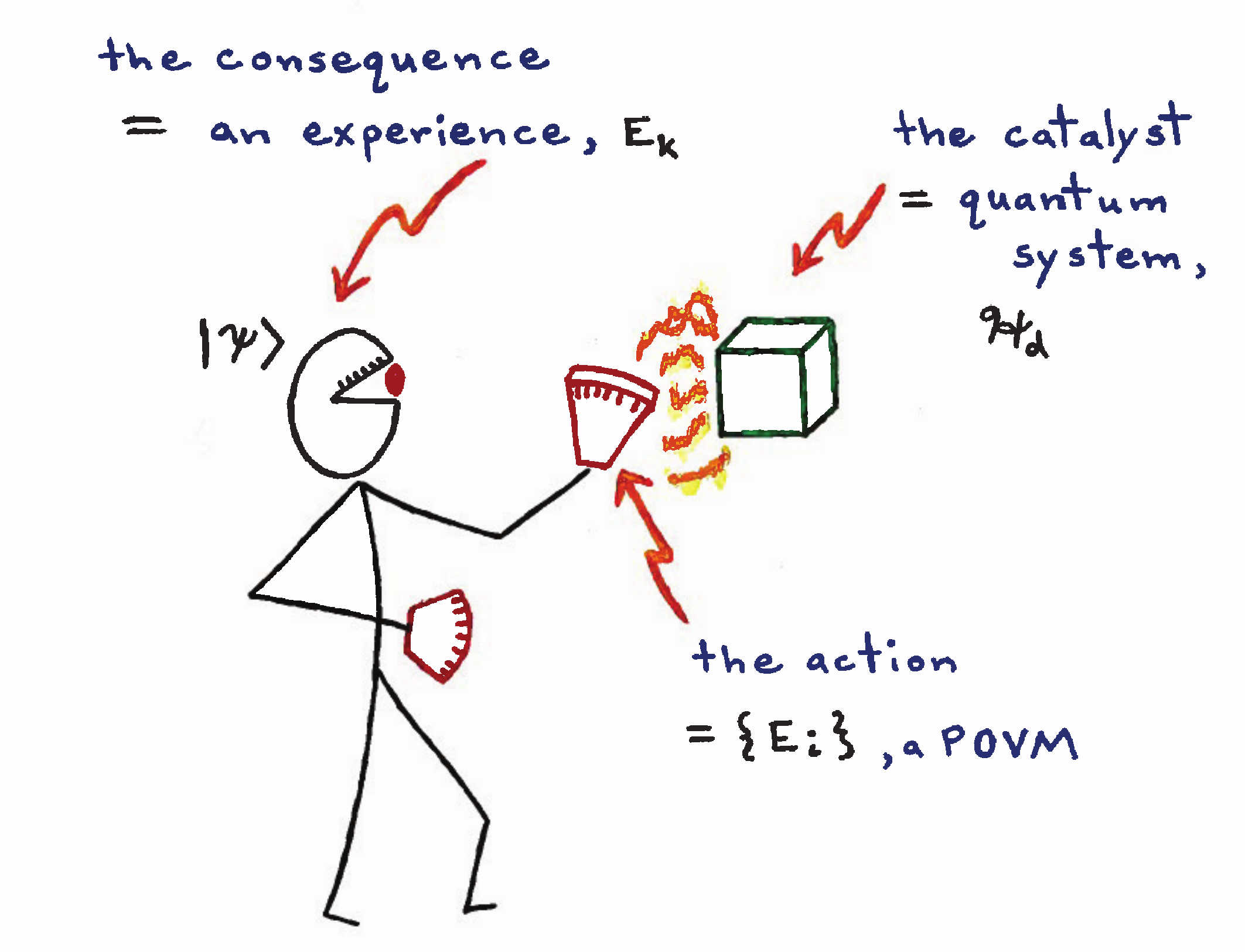}
\end{figure}
\begin{quote} \footnotesize Figure 1: 
The Paulian Idea (Fuchs 2010b) -- in the form of a
  figure inspired by John Archibald Wheeler, whose vision of quantum mechanics has been
  greatly inspiring to us, and overtones of his thought can be found throughout
  our own.  The figure of his that suggested the present one can be found in
  Patton and Wheeler (1975), Wheeler and Patton (1977),
Wheeler (1979),  Wheeler (1980), Wheeler (1982), Wheeler (1994). \newline
 In   contemplating a quantum measurement (though the word is a misnomer), one
  makes a conceptual split in the world: one part is treated as an agent, and
  the other as a kind of reagent or catalyst (one that brings about change in
  the agent itself).  In older terms, the former is an observer and the latter
  a quantum system of some finite dimension $d$.  A quantum measurement
  consists first in the agent taking an {\it action\/} on the quantum system.
  The action is formally captured by some positive operator valued measure $\{E_i\}$(POVM, cf.~footnote 2 below).
The action leads
  generally to an incompletely predictable {\it consequence}, a particular
  personal experience $E_i$ for the agent (Fuchs 2007, 2010a).  The quantum
  state $|\psi\rangle$ makes no appearance but in the agent's head; for it only
  captures his degrees of belief concerning the consequences of his actions,
  and -- in contrast to the quantum system itself -- has no existence in the
  external world.  Measurement devices are depicted as prosthetic hands to make
  it clear that they should be considered an integral part of the agent.  (This
  contrasts with Bohr's view where the measurement device is always treated as
  a classically describable system external to the observer.)  The sparks
  between the measurement-device hand and the quantum system represent the idea
  that the consequence of each quantum measurement is a unique creation within
  the previously existing universe (Fuchs 2010a, 2013). \newline
  Wolfgang Pauli characterized this picture as a ``wider form of the
  reality concept'' than that of Einstein's, which he labeled ``the ideal of
  the detached observer'' (Pauli 1994, Laurikainen 1988, Gieser 2005).
\end{quote}

\newpage

\section{Basic Notions of Quantum Theory \\ from a QBist Point of View} 

Before exploring this further, let us partially formalize in a quick outline
the structure of quantum mechanics from the Bayesian point of view. At the moment we will retain the usual mathematical formulation of the theory, but we will begin the
process of changing the verbal description of what the term ``quantum measurement'' means.

\begin{enumerate}
\vspace{-0.2cm}
\item  Primitive notions: a) the agent, b) things external to the agent, or, more
commonly, ``systems,'' c) the agent's actions on the systems, and d) the
consequences of those actions for her experience.
\vspace{-0.2cm}
\item The formal structure of quantum mechanics is a theory of how the agent ought to
organize her Bayesian probabilities for the consequences of all her potential
actions on the things around her.  Implicit in this is a theory of the
structure of actions.  This works as follows.
\vspace{-0.2cm}
\item When the agent posits a system, she posits a Hilbert space ${\mathcal H}_d$ of dimension $d$ as
the arena for all her considerations.
\vspace{-0.2cm}
\item Actions upon the system are captured by positive-operator valued measures $\{E_i\}$, briefly POVMs,\footnote{See, e.g., Berberian (1966) for a precise definition. In contrast
to a projection-valued measure (projector) with characteristic function \{0,1\} and orthogonal eigenfunctions, 
the characteristic function of a POVM is the entire interval [0,1], and the eigenfunctions are generally not orthogonal. POVMs generalize the idealized idea of quantum measurements as projections and lead to a more realistic picture.}  
on ${\mathcal H}_d$.  Potential consequences of the action are
labeled by the individual elements $E_i$ within the set.
%\footnote{There is a
%  formal similarity between this and the development by Cox (1961), where
%  ``questions'' are treated as sets, and ``answers'' are treated as elements
%  within the sets.} 
That is,
\vspace{-0.2cm}
$$
{\rm ACTION}=\{E_i\} \quad {\rm and} \quad {\rm CONSEQUENCE}= E_k\;.
$$
\vspace{-0.7cm}
\item Quantum mechanics organizes the agent's beliefs by saying that she should
strive to find a single density operator $\rho$ such that her degrees of belief
will always satisfy 
\vspace{-0.1cm}
\begin{eqnarray}
{\rm Prob} \Big({\rm CONSEQUENCE}\,\Big|\,{\rm ACTION}\Big) &=&
{\rm Prob} \Big(E_k\,\Big|\,\{E_i\}\Big) \nonumber\\ &=& {\rm Trace} \ \rho E_k\,,\nonumber
\end{eqnarray}
\vspace{-0.1cm}
no matter what action $\{E_i\}$ is under consideration.
\vspace{-0.1cm}
\item Unitary time evolution and more general quantum operations (completely positive
maps) do not represent objective underlying dynamics, but rather address the
agent's belief changes accompanying the flow of time, as well as belief changes
consequent upon any actions taken.
\vspace{-0.2cm}
\item When the agent posits {\it two\/} things external to herself, the arena for all
her considerations becomes ${\mathcal H}_{d_1}\otimes{\mathcal H}_{d_2}$.
Actions and consequences now become POVMs on ${\mathcal
  H}_{d_1}\otimes{\mathcal H}_{d_2}$. 
\vspace{-0.2cm}
\item The agent can nonetheless isolate the notion of an action on a single one of
the things alone: These are POVMs of the form $\{E_i\otimes I\}$, and similarly
with the systems reversed $\{I \otimes E_i\}$.
\vspace{-0.2cm}
\item Resolving the consequence of an action on {\it one} of the things may cause
the agent to update her expectations for the consequences of any further
actions she might take on the {\it other} thing.  But for those latter
consequences to come about, she must elicit them through an actual action on
the second system.
\end{enumerate}

\vspace{-0.2cm}
With regard to the present discussion, the main points to note are items
4, 7, 8, and 9.  Regarding our usage of the word ``measurement,'' they say that
one should think of it simply as an {\it action} upon the system of interest.
Actions lead to consequences within the experience of the agent, and that is
what a quantum measurement is.  A quantum measurement finds nothing,
but very much {\it makes} something.

It is a simple linguistic move, but it does crucial work for resetting the debate on
quantum foundations. It might indeed have been the case that all this
nonstandard formulation was for nought, turning out to be superfluous.  That
is, though we have spelled out very carefully in item 9 that, ``for those
latter consequences to come about, she must elicit them through an actual
action on the second system'', maybe there would have been nothing wrong in
thinking of the latter (and by analogy the former) quantum measurement as
finding a pre-existing value after all.  But this, we have argued previously
(Caves 2007, Fuchs 2013) would contradict item 8, i.e., that one can isolate a
notion of an action on a single system alone.

Thus, in a QBist painting of quantum mechanics, quantum measurements are
``generative'' in a very real sense.  But by that turn, the consequences of our
actions on physical systems must be egocentric as well.
Measurement outcomes come about for the agent herself. Quantum mechanics is a
single-user theory through and through -- first in the usual Bayesian sense with
regard to personal beliefs, and second in that quantum measurement outcomes are
wholly personal experiences.\footnote{The usual belief otherwise -- for instance
  in Pauli's own formulation (which is ultimately inconsistent with his taking
  measurement devices to be like prosthetic hands), that ``the objective
  character of the description of nature given by quantum mechanics [is]
  adequately guaranteed by the circumstance that \ldots\ the results of
  observation, {\it which can be checked by anyone}, cannot be influenced by
  the observer, once he has chosen his experimental arrangement'' (Pauli 1956, italics
    ours, to pinpoint the offending portion of the formulation) -- we state
  for completeness, is the ultimate source of the Wigner's friend paradox.
  This will be expanded upon in a later work by the authors; for the moment see
  Fuchs (2013).}

Of course, as a single-user theory, quantum mechanics is available to any agent
to guide and better prepare her for her own encounters with the world. And
although quantum mechanics has nothing to say about another agent's personal
experiences, agents can communicate and use the information gained from each
other to update their probability assignments. In the spirit of the Paulian
idea, however, querying another agent means taking an action on him. 

Whenever ``I'' encounter a quantum system, and take an action upon it, it catalyzes a
consequence in my experience that my experience could not have foreseen.
Similarly, by a Copernican-style principle, I should assume the same for ``you'':
Whenever you encounter a quantum system, taking an action upon it, it catalyzes
a consequence in your experience.  By one category of thought we are agents,
but by another category of thought we are physical systems.  And when we take
actions upon each other, the category distinctions are symmetrical.  Like with the bistable 
 perception of ambiguous images (e.g., the Rubin vase), the best the eye can do is flit back and forth between the two
formulations.

\section{The World View of QBism}

The previous paragraphs should have made clear that viewing quantum mechanics
as a single-user theory does not mean there is only one user. QBism does not lead
to solipsism.  Any charge of solipsism is further refuted by two points central
to the Paulian idea (Fuchs 2002b).  One is the conceptual split of the world
into two parts -- one an agent and the other an external quantum system -- that
gets the discussion of quantum measurement off the ground in the first place.
If such a split were not needed for making sense of the question of actions
(actions upon what?\ in what?\ with respect to what?), it would not have
been made. Imagining a quantum measurement without an autonomous quantum system
participating in the process would be as paradoxical as the Zen koan of the
sound of a single hand clapping.  

The second point is that once the agent
chooses an action $\{E_i\}$, the particular consequence $E_k$ of it is beyond
his control.  That is to say, the particular outcome of a quantum measurement
is not a product of his desires, whims, or fancies -- this is the very reason he
uses the calculus of probabilities in the first place: they quantify his
uncertainty (Lindley 2006), an uncertainty that, try as he might, he cannot
get around.  So, implicit in this whole picture -- this whole Paulian idea -- is
an ``external world \ldots\ made of {\it something},'' just as Gardner
calls for.  It is only that quantum theory is a rather small theory: Its
boundaries are set by being a handbook for agents immersed within that ``world
made of {\it something}''.

But a small theory can still have grand import, and quantum mechanics most
certainly does.  This is because it tells us how a user of the theory sees his
role in the world.  Even if quantum mechanics -- viewed as an addition to
probability theory -- is not a theory of the world itself, it is certainly
conditioned by the particular character of this world. Its empirical content
is exemplified by the Born rule, item 5 in the above list,
which takes a specific form rather than an infinity of other possibilities.
Even though quantum theory is now understood as a theory of acts, decisions,
and consequences (Savage 1954), it tells us, in code, about the character of
our particular world.
Apparently, the world is made of a stuff that does not have
``consequences'' waiting around to fulfill our ``actions'' -- it is a world in
which the consequences are generated on the fly.  When we on the inside prod
that stuff on the outside, the world comes to something that neither side could
have foretold.

Indeed, one starts to get a sense of a world picture that is part
personal -- {\it truly} personal -- and part the joint product of all that
interacts.  It is almost as if one can hear in the very formulation of the Born
rule one of William James' many lectures on chance and indeterminism.  Here is
one example (James 1956a): 
\begin{quote} \small
 [Chance] is a purely negative and relative
term, giving us no information about that of which it is predicated, except
that it happens to be disconnected with something else -- not controlled,
secured, or necessitated by other things in advance of its own actual
presence. \ldots\ What I say is that it tells us nothing about what a thing may
be in itself to call it ``chance.'' \ldots\ All you mean by calling it
``chance'' is that this is not guaranteed, that it may also fall out
otherwise. For the system of other things has no positive hold on the
chance-thing. Its origin is in a certain fashion negative: it escapes, and
says, Hands off!\ coming, when it comes, as a free gift, or not at all. \\
This negativeness, however, and this opacity of the chance-thing when thus
considered {\it ab extra}, or from the point of view of previous things or
distant things, do not preclude its having any amount of positiveness and
luminosity from within, and at its own place and moment. All that its
chance-character asserts about it is that there is something in it really of
its own, something that is not the unconditional property of the whole. If the
whole wants this property, the whole must wait till it can get it, if it be a
matter of chance. That the universe may actually be a sort of joint-stock
society of this sort, in which the sharers have both limited liabilities and
limited powers, is of course a simple and conceivable notion.
\end{quote}
And here is another (James 1956b):
\begin{quote} \small
Why may not the world be a sort of
republican banquet of this sort, where all the qualities of being respect one
another's personal sacredness, yet sit at the common table of space and time? \\
To me this view seems deeply probable.  Things cohere, but the act of cohesion
itself implies but few conditions, and leaves the rest of their qualifications
indeterminate.  As the first three notes of a tune comport many endings, all
melodious, but the tune is not named till a particular ending has actually
come, -- so the parts actually known of the universe may comport many ideally
possible complements.  But as the facts are not the complements, so the
knowledge of the one is not the knowledge of the other in anything but the few
necessary elements of which all must partake in order to be together at all.
Why, if one act of knowledge could from one point take in the total
perspective, with all mere possibilities abolished, should there ever have been
anything more than that act?  Why duplicate it by the tedious unrolling, inch
by inch, of the foredone reality?  No answer seems possible. On the other hand,
if we stipulate only a partial community of partially independent powers, we
see perfectly why no one part controls the whole view, but each detail must
come and be actually given, before, in any special sense, it can be said to be
determined at all.  This is the moral view, the view that gives to other powers
the same freedom it would have itself, -- not the ridiculous ``freedom to do
right'', which in my mouth can only mean the freedom to do as {\it I} think
right, but the freedom to do as {\it they} think right, or wrong either. 
\end{quote}

This is a world of ``objective indeterminism'' indeed, but one with no place
for ``objective chance'' in the sense of David Lewis (1986).  From within any part, the future is
undetermined. If one of those parts
is an agent, then it is an agent in a situation of uncertainty.  And where
there is uncertainty, agents should use the calculus of Bayesian probability in
order to make the best go at things.

But we have learned enough from Copernicus to know that egocentrism, whenever
it can be shaken away from a {\it Weltanschauung}, it ought to be.  Whenever
``I'' encounter a quantum system, and take an action upon it, it catalyzes a
consequence in my experience that my experience could not have foreseen.
Similarly, by a Copernican principle, I should assume the same for ``you'':
Whenever you encounter a quantum system, taking an action upon it, it catalyzes
a consequence in your experience.  By one category of thought, we are agents,
but by another category of thought we are physical systems.
And when we take actions upon each other, the category distinctions are symmetrical. 
% Like with ambiguous perception,
%the best the eye can do is flip back and forth between the two
%formulations. 

In the common circles of the philosophy of science there is a
strong popularity in the idea that agentialism can always be reduced to some
complicated property arrived at from physicalism.  But perhaps this
republican-banquet vision of the world that so seems to fit a QBist
understanding of quantum mechanics is telling us that the appropriate ontology
we should seek would treat these dual categories as just that, dual aspects of
a higher, more neutral realm.\footnote{For a few further suggestive things to read in this
  regard, we propose  James (1940, 1996), Lamberth (1999),  Taylor and Wozniak (1996),  Wahl (1925). Neutral monism and dual-aspect monism have become influential frameworks of thinking in contemporary  discussions in the philosophy of mind (cf.~Velmans and Nagasawa 2012).} That is, the concepts
``action'' and ``unforeseen consequence in experience'', both crucial for
clarifying the very meaning of quantum measurement, might just be applicable
after a fashion to arbitrary components of the world -- i.e., venues in which
probability talk has no place.  Understanding or rejecting this idea is the
long road ahead of us.

We leave an old teacher of ours with some closing words that touch on the challenge
William James started us off with:

{\small

\begin{quote}\raggedleft
{\it It is difficult to escape asking a challenging question. Is the
entirety of existence, rather than being built on particles or fields
of force or multidimensional geometry, built upon billions upon
billions of elementary quantum phenomena, those elementary acts of
``observer-participancy'', those most ethereal of all the entities
that have been forced upon us by the progress of science?}

\smallskip

\hspace*{5.4cm} {John Archibald Wheeler 1982}

\end{quote}
}

\section*{References}

\parindent=0mm
\parskip=2mm

Bell J.S.~(1987): {\it Speakable and Unspeakable in Quantum Mechanics}, Cambridge University Press, Cambridge.

Bell J.~(1990): Against measurement. {\it Physics World}, August 1990, 33--41.

Berberian S.K.~(1966): {\it Notes on Spectral Theory}. Van Nostrand,
      New York, p.~5/6.
 
Bernardo J.M.~and Smith A.F.M.~(1994): {\it Bayesian Theory}, Wiley, Chichester.

Caves C.M., Fuchs C.A.~and Schack R.~(2007): Subjective probability and quantum certainty. {\it  Studies in  History and  Philosophy of Modern Physics} {\bf 38}, 255--274.

%Cox R.T.~(1961): {\it The Algebra of Probable Inference}, Johns Hopkins Press, 
%Baltimore.

de Finetti B.~(1990): {\it Theory of Probability, 2 Volumes}, Wiley, New York. Originally published in 1974.

Diaconis R.~and Zabell S.L.~(1982): Updating personal probability. {\it Journal of the American Statistical Association} {\bf 77}, 822--830.

Fuchs C.A.~(2002a): Quantum mechanics as quantum information (and only a little more). Manuscript available at \url{arXiv:quant- ph/0205039v1}. 
Abridged version in {\it Quantum Theory: Reconsideration of Foundations}, ed.~by A.~Khrennikov, V\"axj\"o University Press, V\"axj\"o, pp.~463--543.

Fuchs C.A.~(2002b): The anti-V\"axj\"o interpretation of quantum
mechanics. In {\it Quantum Theory: Reconsideration of Foundations}, ed.~by A.~Khrennikov, V\"axj\"o University Press, V\"axj\"o, pp.~99--116. 
%{\tt arXiv:quant-ph/0204146v1}
%(2002).

Fuchs C.A.~(2007): Delirium quantum: Or, where I will take quantum mechanics if it will let me. In {\sl Foundations of Probability and
    Physics -- 4}, ed.~by G.~Adenier, C.A.~Fuchs, and A.~Yu.~Khrennikov, American Institute of Physics, Melville, pp.~438--462. 
%{\tt arXiv:0906.1968v1 [quant-ph]}. %(2009).

Fuchs C.A.~(2010a): QBism, the perimeter of quantum Bayesianism, accessible at \url{ arXiv:1003.5209}. 

Fuchs C.A.~(2010b): {\it Coming of Age with Quantum Information: Notes on a Paulian Idea}, Cambridge University Press, Cambridge.

Fuchs C.A.~(2013): {\it My Struggles with the Block Universe: Selected
  Correspondence, January 2001--May 2011}, {\tt arXiv:1405.2390}.

Fuchs C.A.~and Schack R.~(2004): Unknown quantum states and operations, a Bayesian view. In {\it Quantum Estimation Theory}, ed.~by M.G.A.~Paris and J.~\v{R}eh\'a\v{c}ek, Springer, Berlin, pp.~151--190.

Gardner M.~(1983): Why I am not a solipsist. In {\it The Whys of a Philosophical Scrivener}, W.~Morrow, New York, pp.~11--31.

Gieser S.~(2005): {\it The Innermost Kernel: Depth Psychology and Quantum Physics}, Springer, Berlin.

Hampton J.M., Moore P.G.~and Thomas H.~(1973): Subjective probability and its measurement. {\it Journal of the Royal Statistical Society Series A} {\bf 136}(1), 21--42.

%Harper W., Chow S.J.~and Murray G.~(2009): Bayesian chance. {\it Synthese} {\bf186}, 
%447--474.

James W.~(1922): {\it Pragmatism, a New Name for Some Old Ways of Thinking:
Popular Lectures on Philosophy}, Longmans, Green and Co., New York.

James W.~(1940): {\it Some Problems of Philosophy}, Longmans, Green and Co., London.

James W.~(1956a): The dilemma of determinism. In {\it The Will to Believe and Other Essays in Popular Philosophy}, Dover, New York, pp.~145--183.

James W.~(1956b): On some Hegelisms. In {\it The Will to Believe and Other Essays in Popular Philosophy}, Dover, New York, pp.~263--298.

James W.~(1988): The many and the one 1903--1904. In {\it Manuscript Essays and Notes}, ed.~by I.K.~Skrupskelis, Harvard University Press, Cambridge.

James W.~(1996): {\it Essays in Radical Empiricism}, University of Nebraska Press, Lincoln.

Jeffrey R.~(2004): {\it Subjective Probability: The Real Thing}, Cambridge University Press, Cambridge.

 Lamberth D.C.~(1999): {\it William James and the Metaphysics of Experience}, Cambridge University Press, Cambridge.

Laurikainen K.V.~(1988): {\it Beyond the Atom: The Philosophical Thought of Wolfgang Pauli}, Springer, Berlin.

Lewis D.~(1986): A subjectivist's guide to objective chance. In {\it Studies in Inductive Logic and Probability, Vol.~II}, Oxford University
Press, Oxford, pp.~83--112.

Lindley D.V.~(1982): Comment on A.P.~Dawid's ``The well-calibrated Bayesian''. {\it Journal of the American Statistical Association} {\bf 77}, 604--613.

Lindley D.V.~(2006): {\sl Understanding Uncertainty}, Wiley-Interscience, Hoboken.

%Mermin N.D.~{2006}: In praise of measurement. {\it Quantum 
%Information Processing} {\bf 5}(4), 239--260.

Patton C.M.~and Wheeler J. A.~(1975): Is physics legislated by cosmogony? In {\sl Quantum Gravity: An Oxford Symposium}, ed.~by C.J.~Isham, R.~Penrose, and D.W.~Sciama, Clarendon  Press, Oxford, pp.~538--605.

Pauli W.~(1956):  Relativit\"atstheorie und Wissenschaft. {\it Helvetica Physica Acta, Supp.} {\bf IV}, 282--286. Reprinted as ``The Theory of Relativity and Science'', in W.~Pauli: {\it  
Writings on Physics and Philosophy}, ed.~by C.P.~Enz and K.~von Meyenn, Springer, Berlin 1994, pp.~107--111.

Peres A.~(1978): Unperformed experiments have no results. {\it American Journal of Physics}, {\bf 46}, 745--747.

Savage L.J.~(1954): {\it The Foundations of Statistics}, Wiley, New York.

Skyrms B.~(1987): Dynamic coherence and probability kinematics. {\it Philosophy of Science} {\bf 54}, 1--20.

Taylor E.~and Wozniak R.H., eds.~(1996): {\it Pure Experience: The Response to William James}, Thoemmes Press, Bristol.

Velmans M.~and Nagasawa Y., eds.~(2012): Monist alternatives to physicalism. Special Issue of the {\it Journal of Consciousness Studies} {\bf 19}(9/10). 

Vineberg S.~(2011): Dutch book arguments. In {\it Stanford Encyclopedia of Philosophy}, ed.~by E.N.~Zalta, accessible at \url{plato.stanford.edu/entries/dutch-book/}.

%von Neumann J.~(1932): {\it Mathematische Grundlagen der Quantenmechanik}, Springer, 
%Berlin.

Wahl J.~(1925): {\it The Pluralist Philosophies of England and America}, Open Court, London.

Wheeler J.A.~(1979):  The quantum and the universe. In {\it Relativity, Quanta,
 and Cosmology in the Development of the Scientific Thought of Albert Einstein, Vol.~II}, ed.~by F.~de Finis, Johnson Reprint, New York, pp.~807--825.

Wheeler J.A.~(1980): Beyond the black hole. In {\it Some Strangeness in the Proportion: A Centennial Symposium to Celebrate the
Achievements of Albert Einstein}, ed.~by H.~Woolf, Addison-Wesley, Reading, pp.~341--375.
%discussion pp.~381--386

Wheeler J.A.~(1982): Bohr, Einstein, and the strange lesson of the quantum. In {\it Mind in  Nature: Nobel Conference XVII, Gustavus Adolphus College}, ed.~by R.Q.~Elvee, Harper \& Row, San Francisco, pp.~1--23.
%discussion pp.~23--30, 88--89, 112--113, and 148--149.

Wheeler J.A.~(1994): Time today. In {\it Physical Origins of Time Asymmetry},
  ed.~by J.J.~Halliwell, J.~P\'erez-Mercader, and W.H.~Zurek, Cambridge University Press, Cambridge, pp.~1--29.

Wheeler J.A.~ and Patton  C.M.~(1977): Is physics legislated by cosmogony? In {\it Encyclopedia of Ignorance: Everything You Ever Wanted to Know about the Unknown}, ed.~by 
R.~Duncan and M.~Weston-Smith, Pergamon, Oxford, pp.~19--35.

\end{document}